\documentclass[preprint,showpacs,preprintnumbers]{revtex4}
\usepackage{amsfonts}
\usepackage{amsmath}
\usepackage[dvips]{graphicx}
\usepackage{subfigure}
\usepackage{epsfig}
\usepackage{dcolumn}
\usepackage{bm}
\usepackage{amssymb}

\begin{document}

\title{Natural PQ symmetry in the 3-3-1 model with a minimal scalar sector}
\author{J. C. Montero}
\email{montero@ift.unesp.br}
\affiliation{Instituto de F\'\i sica Te\'orica--Universidade Estadual Paulista \\
R. Dr. Bento Teobaldo Ferraz 271, Barra Funda\\
S\~ao Paulo - SP, 01140-070, Brazil }
\author{B. L. S\'anchez--Vega}
\email{brucesan@ift.unesp.br}
\affiliation{Instituto de F\'\i sica Te\'orica--Universidade Estadual Paulista \\
R. Dr. Bento Teobaldo Ferraz 271, Barra Funda\\
S\~ao Paulo - SP, 01140-070, Brazil }

\begin{abstract}
In the framework of a 3-3-1 model with a minimal scalar sector we make a
detailed study concerning the implementation of the PQ symmetry in order to
solve the strong CP problem. For the original version of the model, with
only two scalar triplets, we show that the entire Lagrangian is invariant
under a PQ-like symmetry but no axion is produced since an $U(1)$ subgroup
remains unbroken. Although in this case the strong CP problem can still be
solved, the solution is largely disfavored since three quark states are left
massless to all orders in perturbation theory. The addition of a third
scalar triplet removes the massless quark states but the resulting axion is
visible. In order to become realistic the model must be extended to account
for massive quarks and invisible axion. We show that the addition of a
scalar singlet together with a $Z_N$ discrete gauge symmetry can
successfully accomplish these tasks and protect the axion field against
quantum gravitational effects. To make sure that the protecting discrete
gauge symmetry is anomaly free we use a discrete version of the
Green--Schwarz mechanism.
\end{abstract}

\pacs{14.80.Va,\,12.60.Cn,\,12.60.Fr }
\maketitle

\section{Introduction}

The standard model (SM) of the elementary particles physics successfully
describes almost all of the phenomenology of the strong, electromagnetic,
and weak interactions. However, from the experimental point of view, the
need to go to physics beyond the standard model comes from the neutrino
masses and mixing, which are required to explain the solar and atmospheric
neutrino data. On the other hand, from the theoretical point of view, the SM
can not be taken as the fundamental theory since some important contemporary
questions, like the number of generations of quarks and leptons, do not have
an answer in its context. Unfortunately we do not know what the physics
beyond the SM should be. A likely scenario is that at the TeV scale physics
will be described by models which, at least, give some insight into the
unanswered questions of the SM.

A way of introducing new physics is to enlarge the symmetry gauge group. For
example, the gauge symmetry may be $SU(3)_{C}\otimes SU(3)_{L}\otimes
U(1)_{X}$, instead of that of the SM. Models based on this gauge group have
become known as 3-3-1 models~\cite{pisano1992, frampton1992, foot1994}.
Although the 3-3-1 models coincide with the SM at low energies, they explain
some fundamental questions. This is the case of the number of generations
cited above. In the 3-3-1 model framework, the number of generations must be
three, or multiple of three, in order to cancel anomalies. This is because
the model is anomaly-free only if there is an equal number of triplets and
antitriplets, including the color degrees of freedom. In this case, each
generation is anomalous. The anomaly cancelation only occurs for the three,
or multiply of three, generations together, and not generation by generation
like in the SM. This provides, at least, a first step towards the
understanding of the flavor question. Other interesting features of the
3-3-1 models concern the electric charge quantization and the vectorial
character of the electromagnetic interaction. These questions can be
accommodated in the SM. However, in the 3-3-1 models these questions are
related one to another and are independent of the nature of the neutrinos.

In recent literature we find studies about the most different aspects of the
3-3-1 model phenomenology. Among others, a fundamental puzzling aspect is:
Why is the CP nonconservation in the strong interactions so small~\cite
{Pal1995, alex2004}? The last question, quantified by the $\overline{\theta }
$ parameter of the effective QCD\ Lagrangian, is known as the strong CP
problem. Several solutions based on different ideas have been proposed.
According to the framework, they are based on: unconventional dynamics~\cite
{shaposhnikov1988}, spontaneously broken CP~\cite{barr1984, barr1986,
nelson1984}, and an additional chiral symmetry. In the framework of
introducing an additional chiral symmetry, two suggestions have been made.
If this symmetry is not broken, the symmetry is realized in the Wigner-Weyl
manner and the only possible way of relating this unbroken chiral symmetry
with flavor conserving gluons is to have at least one massless quark~\cite
{kaplan1986}. This suggestion is disfavored by standard current algebra
analysis~\cite{gasser1982, Nelson}. The second possibility is that the
global $U\left( 1\right) $ chiral symmetry, known as $U\left( 1\right) _{
\text{PQ}}$~\cite{peccei1997, peccei1977}, is spontaneously broken down
which implies a Nambu-Goldstone boson (NG boson), currently known as the
axion~\cite{kim1979, dine1981, zhitnitsky1980}.

In this paper we consider the strong CP problem in the framework of a
version of the 3-3-1 model in which the scalar sector is minimal~\cite{Long}
. This model has become known as the \textquotedblleft economical 3-3-1
model\textquotedblright . The appealing feature of this 3-3-1 model is the
natural existence of a PQ-like $U\left( 1\right) $ symmetry. To study the
consequences of this symmetry into this model, we organize this paper as
follows: in Sec. II we briefly describe the model, in Sec. III we analyze
the consequences of the natural PQ-like symmetry into the model and find
that the symmetry is realized in the Wigner-Weyl manner implying three
massless quarks, what is in disagree with the standard current algebra
analysis. Thus, we propose the introduction of two new scalar fields, $\eta $
and $\phi $, in order to both give a solution to the massless quarks and
implement the PQ mechanism. Since this mechanism need that the $U(1)_{\text{
PQ}}$ be anomalous in order to solve the strong CP problem, it seems not
\textit{natural} to impose this symmetry to the Lagrangian. However, it
could be understood as being \textit{natural} if it is a residual symmetry
of a larger one which is not anomalous and spontaneously broken. Then, we
consider a $Z_{N}$ discrete gauge symmetry to be a symmetry of the
Lagrangian. The discrete gauge anomalies are canceled by a discrete version
of the Green-Schwarz mechanism.\ After, two $Z_{N}$ symmetries, $Z_{10}$ and
$Z_{11}$, which protect the axion against quantum gravity effects, are
explicitly shown. Finally, our conclusions are given in Sec. IV.

\section{A brief review of the economical 3-3-1 model}

The different models based on a 3-3-1 gauge symmetry can be classified
according to the electric charge operator
\begin{equation}
Q=T^{3}-bT^{8}+X\text{,}
\end{equation}
where $T^{3}$ and $T^{8}$ are the diagonal Gell-Mann matrices, $X$ refers to
the quantum number of the $U\left( 1\right) _{X}$ group, and $b=1/\sqrt{3}$,
$\sqrt{3}$. The embedding $b$ parameter defines the model. Here, we will
consider the model with both $b=1/\sqrt{3}$ and the simplest scalar sector,
which was proposed for the first time in Ref.~\cite{ponce}. It has become
known in the literature as \textquotedblleft economical 3-3-1
model\textquotedblright . This model had origin in a systematic study of all
possible 3-3-1 models without exotic electric charges~\cite{ponce2002}.

To give a brief review of the main features of this model, let us say that
it has a fermionic matter content given by
\begin{align}
\Psi _{aL}& =\left( \nu _{a},e_{a},\left( \nu _{aR}\right) ^{C}\right)
_{L}^{T}\sim \left( \mathbf{1},\mathbf{3},-1/3\right) \text{,}\quad
e_{aR}\sim \left( \mathbf{1},\mathbf{1},-1\right) \text{,}  \notag \\
Q_{\alpha L}& =\left( d_{\alpha },u_{\alpha },d_{\alpha }^{\prime }\right)
_{L}^{T}\sim \left( \mathbf{3},\mathbf{3}^{\ast },0\right) \text{,}\quad
Q_{3L}=\left( u_{3},d_{3},u_{3}^{\prime }\right) _{L}^{T}\sim \left( \mathbf{
\ \ 3},\mathbf{3},1/3\right) \text{,}  \notag \\
u_{aR}& \sim \left( \mathbf{3},\mathbf{1},2/3\right) \text{,}\quad
u_{3R}^{\prime }\sim \left( \mathbf{3},\mathbf{1},2/3\right) \text{, }
\notag \\
d_{aR}& \sim (\mathbf{3},\mathbf{1},-1/3)\text{,}\quad d_{\alpha R}^{\prime
}\sim \left( \mathbf{3},\mathbf{1},-1/3\right) \text{,}
\end{align}
where $a=1$, $2$, $3$, $\alpha =1$, $2$ (from now on Latin and Greek letters
always take the values $1$, $2$, $3$ and $1$, $2$, respectively) and the
values in the parentheses denote quantum numbers based on the $\left(
SU\left( 3\right) _{C},SU\left( 3\right) _{L},U\left( 1\right) _{X}\right) $
factor, respectively. In this model the electric charges of the exotic
quarks are the same as the usual ones, i.e. $Q\left( d_{\alpha }^{\prime
}\right) =-1/3$ and $Q\left( u_{3}^{\prime }\right) =2/3$.

In the bosonic matter content there are only two scalar triplets, $\chi $
and $\rho $
\begin{equation}
\chi =\left( \chi ^{0},\chi ^{-},\chi _{1}^{0}\right) ^{T}\sim \left(
\mathbf{1},\mathbf{3},-1/3\right) \text{,}\quad \rho =\left( \rho ^{+},\text{
}\rho ^{0},\text{ }\rho _{1}^{+}\right) \sim \left( \mathbf{1},\mathbf{3},
2/3\right) \text{.}
\end{equation}
These two scalar broken down spontaneously the $SU\left( 3\right)
_{L}\otimes U\left( 1\right) _{X}$ gauge group. The vacuum expection values,
vevs, in this model satisfy the constraint
\begin{equation*}
V_{\rho ^{0}}\equiv \left\langle \text{Re\thinspace }\rho ^{0}\right\rangle
\text{, }V_{\chi ^{0}}\equiv \left\langle \text{Re\thinspace }\chi
^{0}\right\rangle \ll V_{\chi _{1}^{0}}\equiv \left\langle \text{
Re\thinspace }\chi _{1}^{0}\right\rangle \text{.}
\end{equation*}
With the quark, lepton and scalar multiplets above we have the Yukawa
interactions
\begin{align}
\mathcal{L}_{\text{Y}}^{l}& =Y_{ab}\overline{\Psi _{aL}}e_{bR}\rho
+Y_{ab}^{\prime }\epsilon ^{ijk}\left( \overline{\Psi _{aL}}\right)
_{i}\left( \Psi _{bL}\right) _{j}^{C}\left( \rho ^{\ast }\right) _{k}+\text{
H.c.,}  \label{lag yuk leptons} \\
\mathcal{L}_{\text{Y}}^{q}& =G^{1}\overline{Q_{3L}}u_{3R}^{\prime }\chi
+G_{\alpha \beta }^{2}\overline{Q_{\alpha L}}d_{\beta R}^{\prime }\chi
^{\ast }+G_{a}^{3}\overline{Q_{3L}}d_{aR}\rho  \notag \\
& +G_{\alpha a}^{4}\overline{Q_{\alpha L}}u_{aR}\rho ^{\ast }+G_{a}^{5}
\overline{Q_{3L}}u_{aR}\chi +G_{\alpha a}^{6}\overline{Q_{\alpha L}}
d_{aR}\chi ^{\ast }  \notag \\
& +G_{\alpha }^{7}\overline{Q_{3L}}d_{\alpha R}^{\prime }\rho +G_{\alpha
}^{8}\overline{Q_{\alpha L}}u_{3R}^{\prime }\rho ^{\ast }+\text{H.c.,}
\label{lag yuk quarks}
\end{align}
for leptons and quarks respectively. $Y_{ab}$ and $G^{i}$ are arbitrary
complex matrices and $Y_{ab}^{\prime }$ is an antisymmetric matrix. We use
the convention that an addition over repeated indices is implied. Notice
that the Yukawa interactions given in Eqs.~(\ref{lag yuk leptons}) and (\ref
{lag yuk quarks}) are the most general allowed by the gauge symmetries.
Here, we follow exactly the Refs.~\cite{Long} and \cite{long2006}, i.e. none
additional symmetries are imposed, contrarily to what is done in the 
Ref.~\cite{ponce2006} where a $Z_{2}$ symmetry is imposed.

The most general scalar potential invariant under the gauge symmetry is
\begin{align}
V_{\text{H}}& =\mu _{\chi }^{2}\chi ^{\dagger }\chi +\mu _{\rho }^{2}\rho
^{\dagger }\rho +\lambda _{1}\left( \chi ^{\dagger }\chi \right)
^{2}+\lambda _{2}\left( \rho ^{\dagger }\rho \right) ^{2}  \notag \\
& +\lambda _{3}\left( \chi ^{\dagger }\chi \right) \left( \rho ^{\dagger
}\rho \right) +\lambda _{4}\left( \chi ^{\dagger }\rho \right) \left( \rho
^{\dagger }\chi \right) \text{.}  \label{potencial model a}
\end{align}
One of the main features of this model is that its scalar sector is the
simplest possible. In principle, this should make the scalar potential
analysis easier. A study of the stability of this scalar potential is
presented in Ref.~\cite{ponce2009}.

\section{$U(1)_{\text{PQ}}$ symmetry in the economical 3-3-1 model}

An $U\left( 1\right) _{\text{PQ}}$ symmetry is global and chiral~\cite
{peccei1997,peccei1977}, i.e. it treats the left- and right-handed parts of
a Dirac field differently. Moreover, it must be both a symmetry of the
entire Lagrangian and valid only at the classical level. In renormalizable
theories, the key ingredient of the $U(1)_{\text{PQ}}$ is that it must be
afflicted by a color anomaly, i.e. its associated current, $j_{\mu }^{\text{
PQ}}$, must obey
\begin{equation}
\partial ^{\mu }j_{\mu }^{\text{PQ}}\supset \frac{Ng^{2}}{16\pi ^{2}}G
\widetilde{G}\text{,}
\end{equation}
being $G\widetilde{G}=\frac{1}{2}\epsilon ^{\mu \nu \sigma \tau }G_{\mu \nu
}^{b}G_{\sigma \tau }^{b}$, and $G_{\mu \nu }^{b}$ is the color field
strength tensor $\left( b=1\text{,}...\right. $ $\left. \text{, }8\right) $.
$N$ must not be zero.

Now, we are going to prove that the economical 3-3-1 model entire Lagrangian
is naturally invariant under an $U(1)_{\text{PQ}}$ symmetry transformation.
To do so, we search for how many $U\left( 1\right) $ symmetries the model
has. First of all, we write the relations that these symmetries must obey in
order to keep the entire Lagrangian invariant. From Eqs.~(\ref{lag yuk
leptons}-\ref{potencial model a}) we obtain the following relations
\begin{align}
-X_{Q_{3}}+X_{u_{3R}^{\prime }}+X_{\chi }& =0\text{,}\qquad
-X_{Q}+X_{d_{R}^{\prime }}-X_{\chi }=0\text{,}  \label{e1} \\
-X_{Q_{3}}+X_{u_{R}}+X_{\chi }& =0\text{,}\qquad -X_{Q}+X_{d_{R}}-X_{\chi
}=0 \text{,} \\
-X_{Q_{3}}+X_{d_{R}}+X_{\rho }& =0\text{,}\qquad -X_{Q}+X_{u_{R}}-X_{\rho
}=0 \text{,} \\
-X_{Q_{3}}+X_{d_{R}^{\prime }}+X_{\rho }& =0\text{,}\qquad
-X_{Q}+X_{u_{3R}^{\prime }}-X_{\rho }=0\text{,} \\
-X_{\Psi }+X_{e_{R}}+X_{\rho }& =0\text{,}\qquad -2X_{\Psi }-X_{\rho }=0
\text{,}  \label{e5}
\end{align}
where the notation $X_{\psi }$ above is to be understood as the $U\left(
1\right) $ charge of the $\psi $ field. Solving the equations above, we find
three independent $U\left( 1\right) $ symmetries. One of these is the $
U\left( 1\right) _{X}$ gauge symmetry. The other two are the usual baryon
number symmetry, $U\left( 1\right) _{B}$, and a chiral symmetry acting on
the quarks, $U\left( 1\right) _{\text{PQ}}$. Thus, the model actually has a
larger symmetry: $SU\left( 3\right) _{C}\otimes SU\left( 3\right)
_{L}\otimes U\left( 1\right) _{X}\otimes U\left( 1\right) _{B}\otimes
U\left( 1\right) _{\text{PQ}}$. The two last symmetries are global. This is
summarized in the Table \ref{table 1}.
\begin{table}[th]
\caption{Assignment of quantum charges in the economical 3-3-1 model.}
\label{table 1}\centering
\begin{tabular}{ccccccccc}
\hline\hline
& \,\, $Q_{\alpha L}$ & \,\, $Q_{3L}$ & \,\, ($u_{aR}$, $u_{3R}^{\prime }$)
& \,\,($d_{aR}$, $d_{\alpha R}^{\prime }$) & \,\, $\Psi _{aL}$ & \,\, $
e_{aR} $ & \,\, $\rho $ & \, $\chi $ \\ \hline
$U\left( 1\right) _{X}$ & $0$ & $1/3$ & $2/3$ & $-1/3$ & $-1/3$ & $-1$ & $
2/3 $ & $-1/3$ \\
$U\left( 1\right) _{B}$ & $1/3$ & $1/3$ & $1/3$ & $1/3$ & $0$ & $0$ & $0$ & $
0$ \\
$U\left( 1\right) _{\text{PQ}}$ & $-1$ & $1$ & $0$ & $0$ & $-1/2$ & $-3/2$ &
$1$ & $1$ \\ \hline\hline
\end{tabular}
\end{table}
We can see that the $U\left( 1\right) _{\text{PQ}}$ chiral symmetry is
afflicted by a color anomaly in the following way
\begin{equation}
A_{\text{PQ}}\varpropto -X_{\rho }-2X_{\chi }=-3\text{,}
\end{equation}
where $A_{\text{PQ}}$ is the coefficient of the $\left[ SU\left( 3\right)
_{C}\right] ^{2}U\left( 1\right) _{\text{PQ}}$ anomaly. Therefore, this
chiral symmetry is a PQ-like symmetry. Also, notice that in this case the $
U\left( 1\right) _{\text{PQ}}$ is an accidental symmetry, i.e. it follows
from the gauge local symmetry plus renormalizability. In other words, the
economical model naturally has a PQ symmetry. The naturalness of the $
U\left( 1\right) _{\text{PQ}}$ in the economical 3-3-1 model is a key point.
In our understanding, since $U\left( 1\right) _{\text{PQ}}$ symmetry is
anomalous its imposition is not sensible in the sense that in the absence of
further constraints on very high energy physics we should expect all
relevant and marginally relevant operators that are forbidden only by this
symmetry to appear in the effective Lagrangian with coefficient of order
one, but if this symmetry follows from some other free anomaly symmetry, in
our case from the gauge symmetry, all terms which violate it are then
irrelevant in the renormalization group sense.

Unfortunately, when $\chi $ and $\rho $ acquire vacuum expectation values,
vevs, different from zero, a subgroup of $U(1)_{X}\otimes U(1)_{\text{PQ}}$
remains unbroken, i.e. the symmetry-breaking pattern is
\begin{align}
& SU\left( 3\right) _{L}\otimes U\left( 1\right) _{X}\otimes U\left(
1\right) _{\text{PQ}}\overset{\langle \chi \rangle }{\longrightarrow }
SU\left( 2\right) _{L}\otimes U\left( 1\right) _{Y}\otimes U\left( 1\right)
_{\text{PQ}}^{\prime }  \notag \\
& \text{ \ \ \ \ \ \ \ \ \ \ \ \ \ \ \ \ \ \ \ \ \ \ \ \ \ \ \ \ \ \ \ \ \ \
\ \ }\overset{\langle \rho \rangle }{\longrightarrow }U\left( 1\right)
_{Q}\otimes U\left( 1\right) _{\text{PQ}}^{\prime \prime }\text{,}
\end{align}
where $U(1)_{Q}$ is the electromagnetic symmetry. The $SU\left( 3\right)
_{C} $ and $U(1)_{B}$ groups have been omitted in the expression above
because these are both unbroken and irrelevant to the current analysis. An
explicit expression of the $U\left( 1\right) _{\text{PQ}}^{\prime }$
symmetry can be easily written as
\begin{equation}
U\left( 1\right) _{\text{PQ}}^{\prime }\equiv U\left( 1\right) _{\text{PQ}
}+3U\left( 1\right) _{X}\text{.}
\end{equation}
Also, note that $U\left( 1\right) _{\text{PQ}}^{\prime }$ and $U\left(
1\right) _{\text{PQ}}^{\prime \prime }$ are PQ-like symmetries because these
are quiral and afflicted by a color anomaly.

As a consequence of the unbroken $U\left( 1\right) _{\text{PQ}}^{\prime
\prime }$ chiral symmetry (i.e. $U\left( 1\right) _{\text{PQ}}^{\prime
\prime }$ is realized in the Wigner-Weyl manner), none axion appears in the
scalar mass spectrum. Instead of that, some quarks remain massless after the
spontaneous symmetry breaking, and these will remain massless to all orders
of perturbation theory.

To show the previously said, we explicitly calculate the mass spectra of
scalars and quarks. First, we calculate the scalar mass spectrum
\begin{align}
m_{H_{1},H_{2}}^{2}& =\text{$\lambda $}_{1}V_{\rho ^{0}}^{2}+\left( V_{\chi
^{0}}^{2}+V_{\chi _{1}^{0}}^{2}\right) \text{$\lambda $}_{2}  \notag \\
& \pm \sqrt{\left( V_{\rho ^{0}}^{2}\text{$\lambda $}_{1}-\left( V_{\chi
^{0}}^{2}+V_{\chi _{1}^{0}}^{2}\right) \text{$\lambda $}_{2}\right)
^{2}+\left( V_{\chi ^{0}}^{2}+V_{\chi _{1}^{0}}^{2}\right) V_{\rho ^{0}}^{2}
\text{$\lambda $}_{3}^{2}}\text{,}  \label{masa higgs} \\
m_{H_{3}^{\pm }}^{2}& =\frac{1}{2}\left( V_{\rho ^{0}}^{2}+V_{\chi
^{0}}^{2}+V_{\chi _{1}^{0}}^{2}\right) \text{$\lambda $}_{4}\text{,}
\label{massa higgs cargado}
\end{align}
where $V_{\rho ^{0}}$, $V_{\chi ^{0}}$, $V_{\chi _{1}^{0}}$ are the vevs of $
\rho ^{0}$, $\chi ^{0}$, $\chi _{1}^{0}$, respectively. For simplicity, all
the vevs have been assumed to be reals. Additionally, there are exactly 8 NG
bosons that will become the longitudinal components of the\ 8 gauge bosons~
\cite{ponce}. The absence of one physical massless state (or axion) in the
scalar spectrum shows that the $U\left( 1\right) _{\text{PQ}}^{\prime \prime
}$ symmetry remaining unbroken after the spontaneous symmetry breaking.

On the other hand, in the quark spectra, there are three massless states,
one in the up-quark sector and two in the down-quark sector. First, consider
the up quark mass matrix at the tree level which is written as
\begin{equation}
\overline{\mathbf{u}_{L}}M_{u}^{(0)}\mathbf{u}_{R}\equiv \frac{1}{\sqrt{2}}
\overline{\mathbf{u}_{L}}
\begin{bmatrix}
G_{11}^{4}V_{\rho ^{0}} & \text{$G_{12}^{4}V_{\rho ^{0}}$} &
G_{13}^{4}V_{\rho ^{0}} & G_{1}^{8}V_{\rho ^{0}} \\
G_{21}^{4}V_{\rho ^{0}} & G_{22}^{4}V_{\rho ^{0}} & G_{23}^{4}V_{\rho ^{0}}
& G_{2}^{8}V_{\rho ^{0}} \\
G_{1}^{5}V_{\chi ^{0}} & G_{2}^{5}V_{\chi ^{0}} & G_{3}^{5}V_{\chi ^{0}} &
G^{1}V_{\chi ^{0}} \\
G_{1}^{5}V_{\chi _{1}^{0}} & G_{2}^{5}V_{\chi _{1}^{0}} & G_{3}^{5}V_{\chi
_{1}^{0}} & G^{1}V_{\chi _{1}^{0}}
\end{bmatrix}
\mathbf{u}_{R}\text{,}
\end{equation}
where $\overline{\mathbf{u}_{L}}\equiv \left( \overline{u_{1L}},\overline{
u_{2L}},\overline{u_{3L}},\overline{u_{3L}^{\prime }}\right) $ and $\mathbf{
u }_{R}\equiv \left( u_{1R},u_{2R},u_{3R},u_{3R}^{\prime }\right) ^{\text{T}
} $ . The third and fourth rows of the $M_{u}^{(0)}$ matrix are
proportional, thus there is a massless up quark (we call this massless up
quark simply as $u$) at the tree level. An analytical expression for this
massless state can be given but it is useless for our analysis. Later we
give arguments that the $u$ quark remain massless to all orders of
perturbation theory~\cite{banks}. Similarly, the down-quark mass matrix at
the tree level, $M_{d}^{(0)}$, defined as $\frac{1}{\sqrt{2}}\overline{
\mathbf{d}_{L}} M_{d}^{(0)}\mathbf{d}_{R}$, reads
\begin{equation}
\begin{bmatrix}
G_{11}^{6}V_{\chi ^{0}}\text{ \ } & \text{$G_{12}^{6}V_{\chi ^{0}}$ \ } &
G_{13}^{6}V_{\chi ^{0}}\text{ \ } & G_{11}^{2}V_{\chi ^{0}}\text{ \ } &
G_{12}^{2}V_{\chi ^{0}}\text{ \ } \\
G_{21}^{6}V_{\chi ^{0}}\text{ \ } & G_{22}^{6}V_{\chi ^{0}}\text{ \ } &
G_{23}^{6}V_{\chi ^{0}}\text{ \ } & G_{21}^{2}V_{\chi ^{0}}\text{ \ } &
G_{22}^{2}V_{\chi ^{0}}\text{ \ } \\
G_{1}^{3}V_{\rho ^{0}}\text{ \ } & G_{2}^{3}V_{\rho ^{0}}\text{ \ } &
G_{3}^{3}V_{\rho ^{0}}\text{ \ } & G_{1}^{7}V_{\rho ^{0}}\text{ \ } &
G_{2}^{7}V_{\rho ^{0}}\text{ \ } \\
G_{11}^{6}V_{\chi _{1}^{0}}\text{ \ } & G_{12}^{6}V_{\chi _{1}^{0}}\text{ \ }
& G_{13}^{6}V_{\chi _{1}^{0}}\text{ \ } & G_{11}^{2}V_{\chi _{1}^{0}}\text{
\ } & G_{12}^{2}V_{\chi _{1}^{0}}\text{ \ } \\
G_{21}^{6}V_{\chi _{1}^{0}}\text{ \ } & G_{22}^{6}V_{\chi _{1}^{0}}\text{ \ }
& G_{23}^{6}V_{\chi _{1}^{0}}\text{ \ } & G_{21}^{2}V_{\chi _{1}^{0}}\text{
\ } & G_{22}^{2}V_{\chi _{1}^{0}}\text{ \ }
\end{bmatrix}
\text{,}
\end{equation}
where $\overline{\mathbf{d}_{L}}\equiv \left( \overline{d_{1L}},\overline{
d_{2L}},\overline{d_{3L}},\overline{d_{1L}^{\prime }},\overline{
d_{2L}^{\prime }}\right) $ and $\mathbf{d}_{R}\equiv \left(
d_{1R},d_{2R},d_{3R},d_{1R}^{\prime },d_{2R}^{\prime }\right) ^{\text{T}}$.
Since the first and fourth rows, and the second and fifth rows, are
proportional to each other, the $M_{d}^{(0)}$ matrix has two eigenvalues
equal to zero (we call these massless down quarks as $d$ and $s$). Thus, the
economical model has three massless quark states: one in the up-quark sector
and two in the down-quark sector. In other words, the economical 3-3-1 model
has a remaining unbroken quiral symmetry, $U\left( 1\right) _{\text{PQ}
}^{\prime \prime }$ that allows to transform $u_{L}\rightarrow e^{i\alpha
}u_{L}$, $d_{L}\rightarrow e^{i\alpha }d_{L}$, $s_{L}\rightarrow e^{i\alpha
\gamma }s_{L}$, leaving the Lagrangian invariant. This symmetry will protect
these massless quarks to acquire mass at any level of perturbation theory~
\cite{banks}. At this point it is important to say that, since the $U\left(
1\right) _{\text{PQ}}^{\prime \prime }$ symmetry is anomalous, these quarks
will acquire mass only through QCD non-perturbative effects (for example, by
instanton effects~\cite{Hooft}). Although, the quarks could acquire some
mass through these non-perturbative processes, this is in conflict with both
chiral QCD and lattice calculation where the ratio $m_{u}/m_{d}$ is $
0.410\pm 0.036$ \cite{gasser1982, Nelson, kim2009}.

Before considering a possible solution to the problem mentioned above, for
the sake of completeness, we find important to say that in the Ref.~\cite
{Long} one-loop contributions to the up-quark mass matrix were calculated,
even though a subtle flaw makes these contributions no right. To demonstrate
that, we exactly follow the same lines of the Ref.~\cite{Long}. There, in
the section IV, the authors consider for simplicity, one-loop contributions
to the sub-matrix
\begin{equation}
M_{u_{3}u_{3}^{\prime }}^{(0)}\equiv \frac{1}{\sqrt{2}}
\begin{bmatrix}
G_{3}^{5}V_{\chi ^{0}} & \text{ \ \ }G^{1}V_{\chi ^{0}} \\
G_{3}^{5}V_{\chi _{1}^{0}} & \text{ \ \ }G^{1}V_{\chi _{1}^{0}}
\end{bmatrix}
\text{,}  \label{matriz u}
\end{equation}
where $M_{u_{3}u_{3}^{\prime }}^{(0)}$ is written in the base $\left( u_{3}
\text{, }u_{3}^{\prime }\right) $. The other two massive quark states, $
u_{1} $ and $u_{2}$, which acquire mass at tree level ($m_{1}=G_{11}^{4}V_{
\rho ^{0}}/\sqrt{2}$, $m_{2}=G_{22}^{4}V_{\rho ^{0}}/\sqrt{2}$, see Eq.(27)
in Ref.~\cite{Long}) are not important in the analysis. The matrix Eq.~(\ref
{matriz u}) mixes together the states $u_{3}$ and $u_{3}^{\prime }$. A
combination of them will be a massless quark and the orthogonal combination
acquires a mass $\sim V_{\chi _{1}^{0}}$.

\begin{figure}[h]
\center
\subfigure[ref1][]{\includegraphics[width=0.4\textwidth]{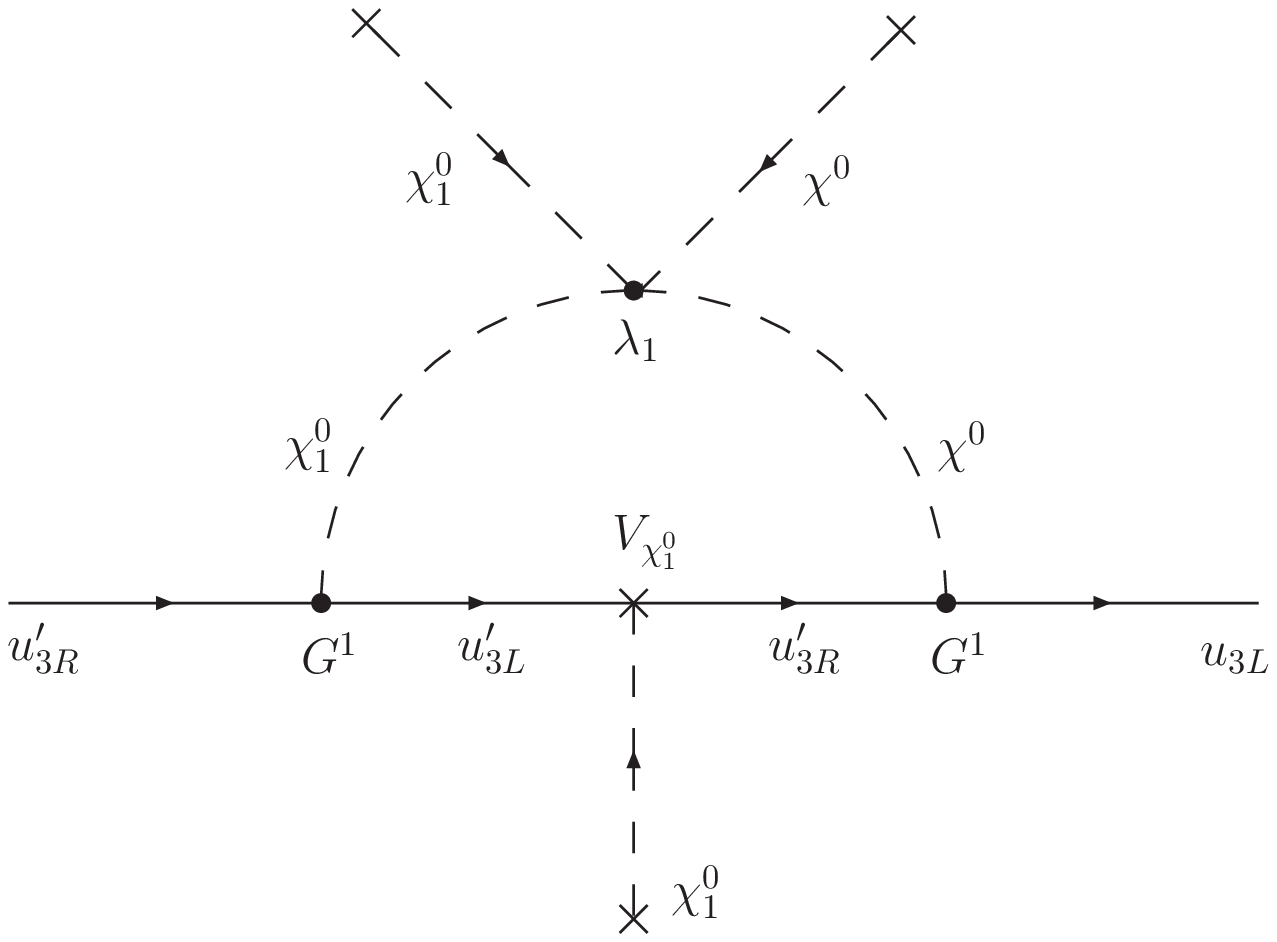}}
\label{a} \qquad
\subfigure[ref2][]{\includegraphics[width=0.4
\textwidth]{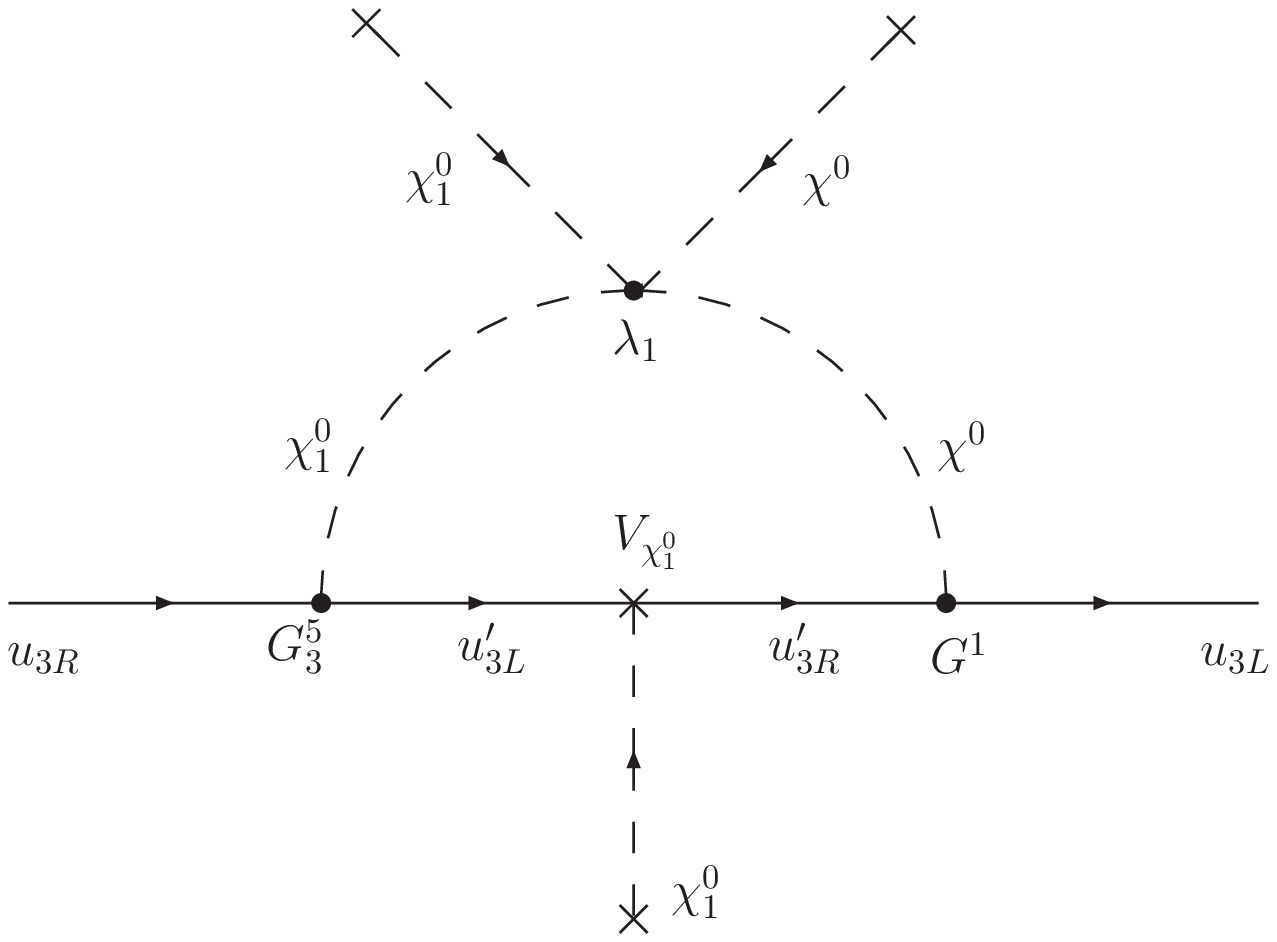}} \newline
\subfigure[ref3][]{\includegraphics[width=0.4\textwidth]{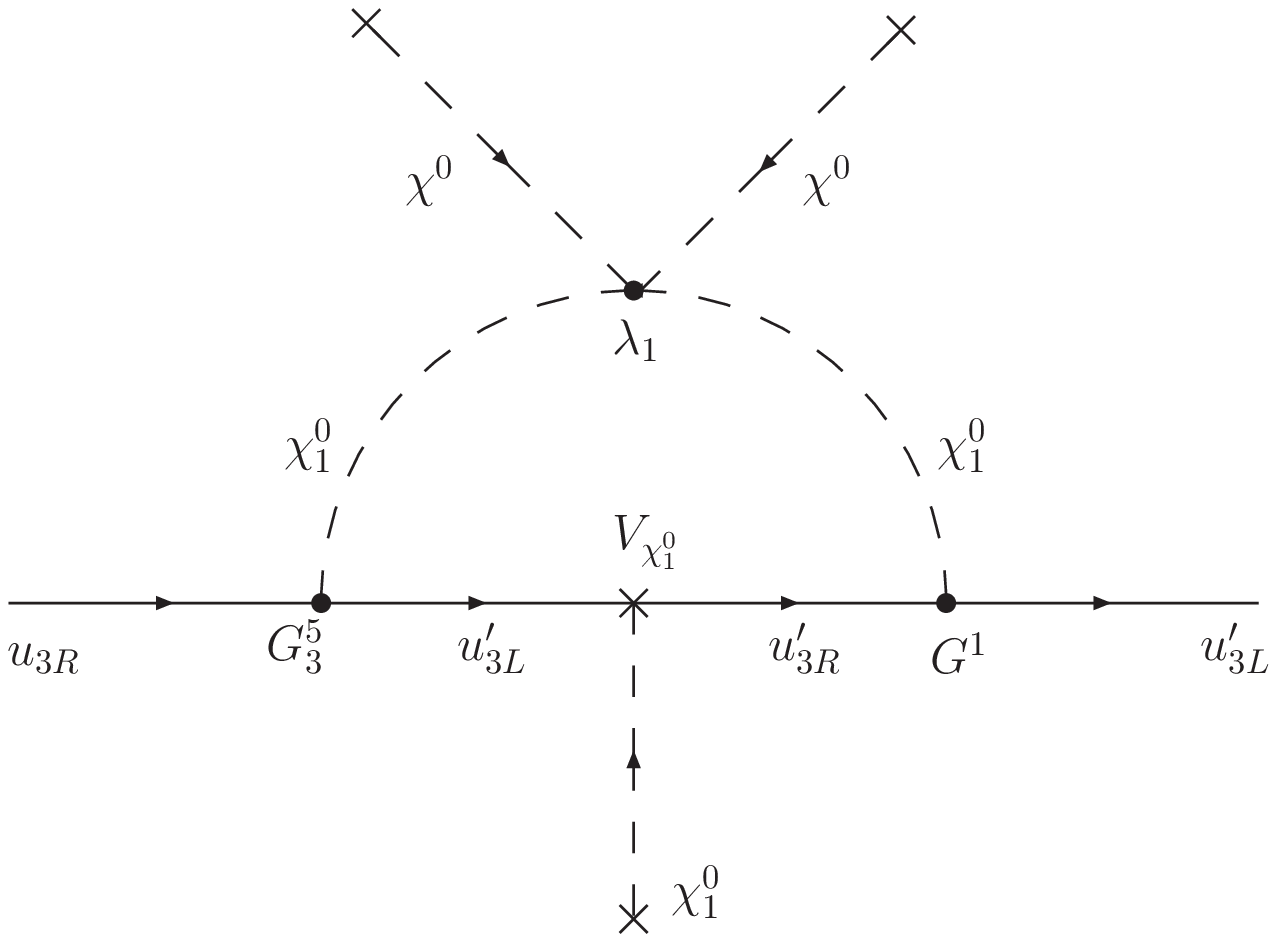}}
\qquad
\subfigure[ref4][]{\includegraphics[width=0.4
\textwidth]{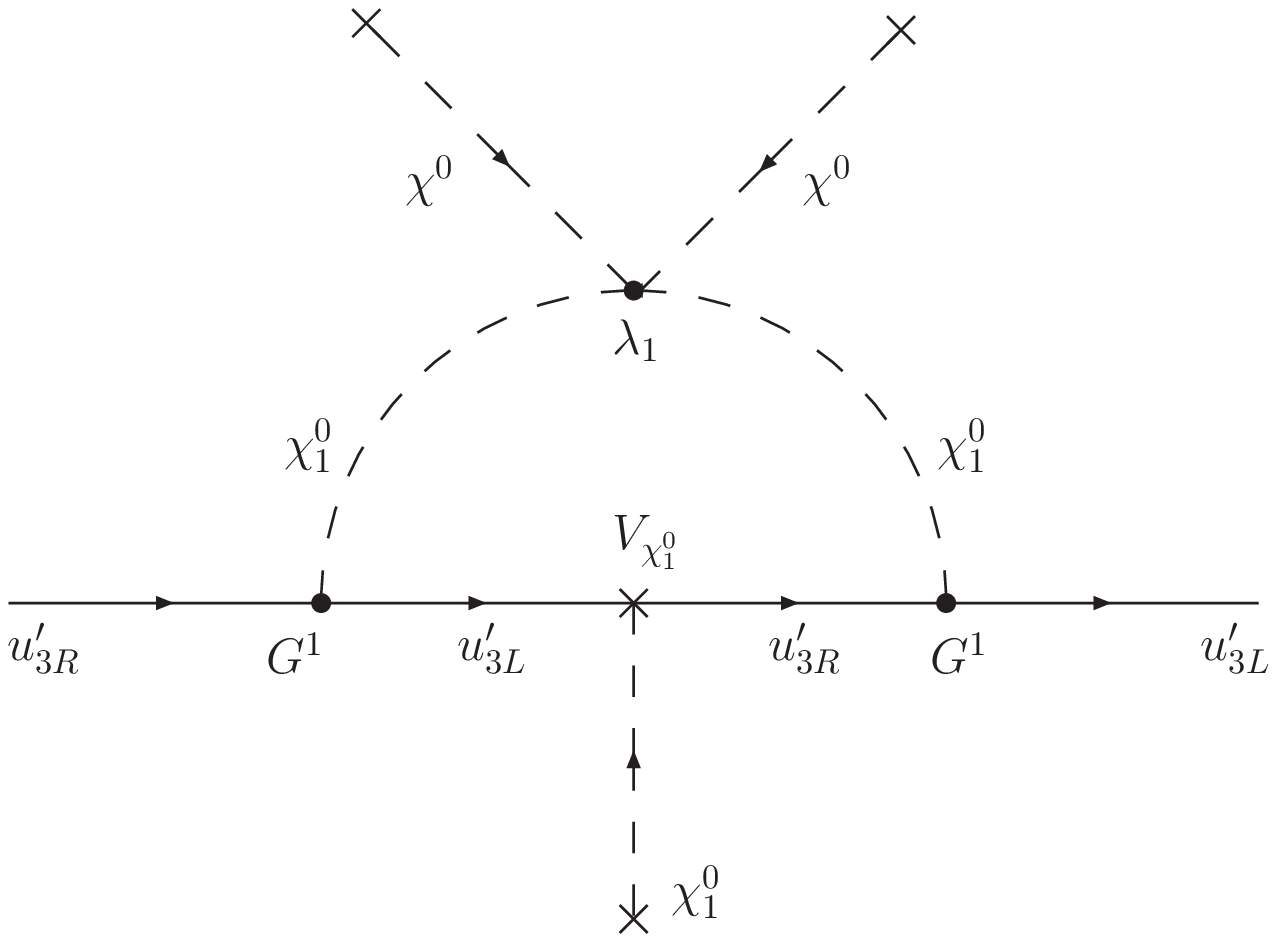}}
\caption{\textit{One-loop contributions to the up-quark mass matrix.}}
\label{one loop}
\end{figure}

Now, the idea is to calculate the one-loop contributions coming from the
Feynman diagrams in the Fig.~\ref{one loop} to the up-quark mass sub-matrix
defined in Eq.~(\ref{matriz u}).
Following the Ref.~\cite{Long}, we get
\begin{align}
\Delta _{u_{3L}\text{,}u_{3R}^{\prime }}& =-2iV_{\chi ^{0}}V_{\chi
_{1}^{0}}\lambda _{1}M_{u_{3}^{\prime }}\left( G^{1}\right) ^{2}  \notag \\
& \times \int \frac{d^{4}p}{\left( 2\pi \right) ^{4}}\frac{p^{2}}{\left(
p^{2}-M_{u_{3}^{\prime }}^{2}\right) ^{2}\left( p^{2}-M_{\chi
^{0}}^{2}\right) \left( p^{2}-M_{\chi _{1}^{0}}^{2}\right) }  \notag \\
& \equiv 2V_{\chi ^{0}}V_{\chi _{1}^{0}}\lambda _{1}M_{u_{3}^{\prime
}}\left( G^{1}\right) ^{2}I\left( M_{u_{3}^{\prime }}^{2}\text{, }M_{\chi
^{0}}^{2}\text{, }M_{\chi _{1}^{0}}^{2}\right) \text{,}
\end{align}
where $I\left( M_{u_{3}^{\prime }}^{2}\text{, }M_{\chi ^{0}}^{2}\text{, }
M_{\chi _{1}^{0}}^{2}\right) $ is defined as
\begin{equation}
I\left( M_{u_{3}^{\prime }}^{2}\text{, }M_{\chi ^{0}}^{2}\text{, }M_{\chi
_{1}^{0}}^{2}\right) \equiv -i\int \frac{d^{4}p}{\left( 2\pi \right) ^{4}}
\frac{p^{2}}{\left( p^{2}-M_{u_{3}^{\prime }}^{2}\right) ^{2}\left(
p^{2}-M_{\chi ^{0}}^{2}\right) \left( p^{2}-M_{\chi _{1}^{0}}^{2}\right) }
\text{,}  \label{integral}
\end{equation}
and $\Delta _{u_{3L}\text{,}u_{3R}^{\prime }}$is the one-loop contribution
to the element $\left( M_{u_{3}u_{3}^{\prime }}^{(0)}\right) _{12}$ given by
the Feynman diagram (a) in the Fig.~\ref{one loop}. The value of the
integral in Eq.~(\ref{integral}) is not relevant in our analysis and thus it
is not calculated. Now, $\Delta _{u_{3L}\text{,}u_{3R}}$ is found in a
similar way from the diagram (b) in the Fig.~\ref{one loop},
\begin{align}
\Delta _{u_{3L},u_{3R}}& =-2iV_{\chi ^{0}}V_{\chi _{1}^{0}}\lambda
_{1}M_{u_{3}^{\prime }}G_{3}^{5}G^{1}  \notag \\
& \times \int \frac{d^{4}p}{\left( 2\pi \right) ^{4}}\frac{p^{2}}{\left(
p^{2}-M_{u_{3}^{\prime }}^{2}\right) ^{2}\left( p^{2}-M_{\chi
^{0}}^{2}\right) \left( p^{2}-M_{\chi _{1}^{0}}^{2}\right) }  \notag \\
& =\frac{G_{3}^{5}}{G^{1}}\Delta _{u_{3L}\text{,}u_{3R}^{\prime }}\text{.}
\end{align}
One-loop contributions to $\left( M_{u_{3}u_{3}^{\prime }}^{(0)}\right)
_{21} $ and $\left( M_{u_{3}u_{3}^{\prime }}^{(0)}\right) _{22}$, found from
the Feynman diagrams (c) and (d), respectively, are also proportional to
each other, i.e.
\begin{equation}
\Delta _{u_{3L}^{\prime }\text{,}u_{3R}}=\frac{G_{3}^{5}}{G^{1}}\Delta
_{u_{3L}^{\prime }\text{,}u_{3R}^{\prime }}\text{.}
\end{equation}
Therefore, when considering simultaneously all the one-loop contributions
above, the $M_{u_{3}u_{3}^{\prime }}^{(0)}$ becomes
\begin{equation}
\frac{1}{\sqrt{2}}
\begin{bmatrix}
G_{3}^{5}\left( V_{\chi ^{0}}+\frac{\Delta _{u_{3L}\text{,}u_{3R}^{\prime }}
}{G^{1}}\right) & \text{ \ \ }G^{1}\left( V_{\chi ^{0}}+\frac{\Delta
_{u_{3L} \text{,}u_{3R}^{\prime }}}{G^{1}}\right) \\
G_{3}^{5}\left( V_{\chi _{1}^{0}}+\frac{\Delta _{u_{3L}^{\prime }\text{,}
u_{3R}^{\prime }}}{G^{1}}\right) & \text{ \ \ }G^{1}\left( V_{\chi
_{1}^{0}}+ \frac{\Delta _{u_{3L}^{\prime }\text{,}u_{3R}^{\prime }}}{G^{1}}
\right)
\end{bmatrix}
\text{.}
\end{equation}
This matrix still has determinant equal to zero. In other words, we have
shown that one combination of the up quarks still remains massless, as it
should be. In the down quark sector a similar analysis can be easily made.
Thus, what makes the contributions to the up-quark and down-quark masses
made in the Ref.~\cite{Long} not right, is that those contributions were not
considered simultaneously.

To conclude, the 3-3-1 economical model has three massless quarks (one up
quark and two down quarks) to all order of perturbation theory, which is in
conflict with both chiral QCD and lattice calculation where the ratio $
m_{u}/m_{d}$ is $0.410\pm 0.036$ \cite{Nelson}. Therefore, the economical
model is not realistic and it must be modified to overcome that difficulty.
One manner of doing that is introducing a new scalar triplet, $\eta $:
\begin{equation}
\eta =\left( \eta ^{0},\eta ^{-},\eta _{1}^{0}\right) ^{T}\sim \left(
\mathbf{1},\mathbf{3},-1/3\right) \text{.}
\end{equation}
When the scalar triplet, $\eta ,$ is introduced into the model, the Yukawa
Lagrangian given in Eq.~(\ref{lag yuk quarks}) has the following extra terms
\begin{align}
\mathcal{L}_{\text{Y, extra}}^{q}& =G_{a}^{9}\overline{Q_{3L}}u_{aR}\eta
+G_{\alpha a}^{10}\overline{Q_{\alpha L}}d_{aR}\eta ^{\ast }  \notag \\
& +G^{11}\overline{Q_{3L}}u_{3R}^{\prime }\eta +G_{\alpha \beta }^{12}
\overline{Q_{\alpha L}}d_{\beta R}^{\prime }\eta ^{\ast }+\text{H.c.,}
\label{yukawa quark extra}
\end{align}
and the most general scalar potential invariant under the gauge symmetry, $
V=V_{\text{H}}+V_{\text{NH}}$, has now the following extra terms
\begin{align}
V_{\text{H, extra}}& =\mu _{\eta }^{2}\eta ^{\dagger }\eta +\lambda
_{5}\left( \eta ^{\dagger }\eta \right) ^{2}+\eta ^{\dagger }\eta \left[
\lambda _{6}\left( \rho ^{\dagger }\rho \right) +\lambda _{7}\left( \chi
^{\dagger }\chi \right) \right]  \notag \\
& +\lambda _{8}\left( \rho ^{\dagger }\eta \right) \left( \eta ^{\dagger
}\rho \right) +\lambda _{9}\left( \chi ^{\dagger }\eta \right) \left( \eta
^{\dagger }\chi \right) \text{,}  \label{potencial hermitiano}
\end{align}
and
\begin{align}
V_{\text{NH}}& =\mu _{4}^{2}\chi ^{\dagger }\eta +f\epsilon ^{ijk}\eta
_{i}\rho _{j}\chi _{k}+\lambda _{10}\left( \chi ^{\dagger }\eta \right)
^{2}+\lambda _{11}\left( \chi ^{\dagger }\rho \right) \left( \rho ^{\dagger
}\eta \right)  \notag \\
& +\lambda _{12}\left( \chi ^{\dagger }\eta \right) \left( \eta ^{\dagger
}\eta \right) +\lambda _{13}\left( \chi ^{\dagger }\eta \right) \left( \rho
^{\dagger }\rho \right) +\lambda _{14}\left( \chi ^{\dagger }\eta \right)
\left( \chi ^{\dagger }\chi \right) +\text{H.c. .}
\label{potencial no hermitiano}
\end{align}
Now, when the scalar triplets acquire vevs, it is straightforward to see
that the quark mass matrices do not have determinant equal to zero, thus all
the quarks are massive. Additionally, as we will show below, there will be
none accidental anomalous PQ--like symmetry.

Returning to the question of the PQ symmetry, we note that due to these new
terms in the Lagrangian, the charges of the $U\left( 1\right) $ symmetries
must obey the following relations
\begin{align}
-X_{Q_{3}}+X_{u_{R}}+X_{\eta }& =0\text{,}\qquad -X_{Q_{3}}+X_{u_{R}^{\prime
}}+X_{\eta }=0\text{,}  \label{ea1} \\
-X_{Q}+X_{d_{R}^{\prime }}-X_{\eta }& =0\text{,}\qquad
-X_{Q}+X_{d_{R}}-X_{\eta }=0\text{,} \\
X_{\rho }+X_{\eta }+X_{\chi }& =0\text{,}\qquad -X_{\chi }+X_{\eta }=0\text{
, }  \label{ea3}
\end{align}
besides the ones given in Eqs.~(\ref{e1}-\ref{e5}). Solving Eqs.~(\ref{e1}-
\ref{e5}) and Eqs.~(\ref{ea1}-\ref{ea3}) simultaneously, we find that there
are only two $U\left( 1\right) $ symmetries, $U\left( 1\right) _{X}$ and $
U\left( 1\right) _{B}$. The assignment of quantum charges for these two $
U\left( 1\right) \,$symmetries when $\eta $ is included is shown in the
Table \ref{table 2}.
\begin{table}[th]
\caption{Assignment of quantum charges when $\protect\eta $ is included.}
\label{table 2}\centering
\begin{tabular}{ccccccccc}
\hline\hline
& \,\, $Q_{\alpha L}$ & \,\, $Q_{3L}$ & \,\, ($u_{aR}$, $u_{3R}^{\prime }$)
& \,\, ($d_{aR}$, $d_{\alpha R}^{\prime }$) & \,\,$\Psi _{aL}$ & \,\, $
e_{aR} $ & \,\, $\rho $ & \,\, ($\chi $, $\eta $) \\ \hline
$U\left( 1\right) _{X}$ & $0$ & $1/3$ & $2/3$ & $-1/3$ & $-1/3$ & $-1$ & $
2/3 $ & $-1/3$ \\
$U\left( 1\right) _{B}$ & $1/3$ & $1/3$ & $1/3$ & $1/3$ & $0$ & $0$ & $0$ & $
0$ \\ \hline\hline
\end{tabular}
\end{table}
Thus, in this case, in contrast to the previous one, the $U(1)_{\text{PQ}}$
is not allowed by the gauge symmetry. But, if the Lagrangian is slightly
modified by imposing a $Z_{2}$ symmetry such that $\chi \rightarrow -\chi
,\,\,u_{3R}^{\prime }\rightarrow -u_{3R}^{\prime },\,\,d_{\beta R}^{\prime
}\rightarrow -d_{\beta R}^{\prime }\text{,}$ and all the other fields being
even under $Z_{2}$, the trilinear term of the scalar potential, $f\epsilon
^{ijk}\eta _{i}\rho _{j}\chi _{k}$, is eliminated. Consequently, the $
U\left( 1\right) _{\text{PQ}}$ symmetry is automatically introduced. This
can be seen by solving Eqs.~(\ref{e1}-\ref{e5}) and Eqs.~(\ref{ea1}-\ref{ea3}) 
without the equation
\begin{equation}
X_{\rho }+X_{\eta }+X_{\chi }=0\text{.}
\end{equation}
Note that, in addition to the assignment of quantum charges given in the
Table \ref{table 1}, the charge $U\left( 1\right) _{\text{PQ}}$ of the $\eta
$ triplet scalar is $1$. Unfortunately, the axion that appear when the
neutral components of the scalar triplets acquire vev is visible. This is
easy to see as follows. In this model the $\chi $ field is responsible to
break the symmetry from $SU\left( 3\right) _{C}\otimes SU\left( 3\right)
_{L}\otimes U\left( 1\right) _{X}$ to $SU\left( 3\right) _{C}\otimes
SU\left( 2\right) _{L}\otimes U\left( 1\right) _{Y}$. Thus, for obtaining an
invisible axion, $V_{\chi _{1}^{0}}$ that breaks the PQ symmetry must to be
greater than $10^{9}$ GeV. But, when $\chi $ acquires a vev the combination $
U\left( 1\right) _{\text{PQ}}^{\prime }=U\left( 1\right) _{\text{PQ}
}+3U\left( 1\right) _{X}$ is not broken. Therefore, the new PQ symmetry is
truly broken when the $\rho $ field acquires a vev. As $V_{\rho
^{0}}\lesssim 246$ GeV, the axion induced is visible. A visible axion was
long ago ruled out by experiments \cite{Bardeen}.

One usual way to resolve that problem is to introduce an electroweak scalar
singlet, $\phi $\cite{kim1979, dine1981}. Its role is to break the PQ
symmetry at a scale much larger than the electroweak scale. This field does
not couple directly to quarks and leptons, however, it acquires a PQ charge
by coupling to the scalar triplets. With the PQ charges given in the Table
\ref{table 1}, the $\phi $ scalar acquires a PQ charge by coupling to the $
\eta $, $\rho $, $\chi $ scalar triplets through the interaction term
\begin{equation}
\lambda _{\text{PQ}}\epsilon ^{ijk}\eta _{i}\rho _{j}\chi _{k}\phi \text{.}
\label{acoplamiento pq}
\end{equation}
From this coupling, the $\phi $ field obtain a PQ charge of $-3$. Also,
notice that this term is permitted provided the $\phi $ field is odd under
the $Z_{2}$ symmetry, i.e. $Z_{2}\left( \phi \right) =-\phi $. However, the $
Z_{2}$ and gauge symmetries do not prohibit some terms in the scalar
potential violating the PQ symmetry, such as $\phi ^{2}$, $\phi ^{3}$, $\phi
^{4}$, $\rho ^{\dagger }\rho \phi ^{2}$, $\eta ^{\dagger }\eta \phi ^{2}$, $
\chi ^{\dagger }\chi \phi ^{2}$; from appearing. Thus, the PQ symmetry
should be imposed. Since the PQ symmetry is anomalous, it is some awkward to
do so. However, there is a way to overcome this difficulty. Consider that
the entire Lagrangian is invariant under a $Z_{N}$ discrete gauge symmetry~
\cite{wilczek1989}, with $N\geq 5$, instead of a $Z_{2}$ symmetry. The $
Z_{N} $ charge assignment that allows the scalar potential to be naturally
free of awkward terms violating the PQ symmetry must satisfy the following
minimal conditions
\begin{eqnarray}
Z_{N}\left( \phi \right) &\neq &\left( 0\text{, }N/2\text{, }N/3\text{, }
N/4\right) \text{,}  \label{condiciones minimas1} \\
\text{ }Z_{N}\left( \eta \right) +Z_{N}\left( \rho \right) +Z_{N}\left( \chi
\right) &\neq &pN\text{,}  \label{condiciones minimas2} \\
\text{ }-Z_{N}\left( \chi \right) +Z_{N}\left( \eta \right) &=&rN\text{; \ }p
\text{, }r\in
\mathbb{Z}
\text{,}  \label{condiciones minimas3}
\end{eqnarray}
and, obviously, the other ones that leave the rest of the Lagrangian
invariant under $Z_{N}$. The $-Z_{N}\left( \chi \right) +Z_{N}\left( \eta
\right) =rN$ condition, with $r\in
\mathbb{Z}
$, is necessary to allow the terms in the scalar potential given in Eq.~(\ref
{potencial no hermitiano}), except the trilinear $f\epsilon ^{ijk}\eta
_{i}\rho _{j}\chi _{k}$ term, and thus, avoid the appearance of an
additional dangerous massless scalar in the physical spectrum. In other
words, with the conditions imposed by\ Eqs.~(\ref{condiciones minimas1}-\ref
{condiciones minimas3}) for this $Z_{N}$ discrete symmetry, none of
Lagrangian terms, except the violating PQ terms, such as $f\epsilon
^{ijk}\eta _{i}\rho _{j}\chi _{k}$, $\phi ^{2}$, $\phi ^{3}$, $\phi ^{4}$,
etc; are prohibit from appearing.

Furthermore, to stabilize the axion solution from quantum gravitational
effects~\cite{marc1992, holman1992} we will make use of the $Z_{N}$ discrete
symmetry with anomaly cancelation by a discrete version of the
Green-Schwarz mechanism~\cite{green1984,green1985,green1985b, babu2003}.
Quantum gravity effective operators, allowed by the gauge symmetry, of the
form $\phi ^{N}/M_{\text{Pl}}^{N-4}$ can induce a non-zero $\overline{\theta
}$ given by
\begin{equation}
\overline{\theta }\simeq \frac{f_{a}^{N}}{\Lambda _{\text{QCD}}^{4}M_{\text{
Pl}}^{N-4}}\text{.}
\end{equation}
From the neutron electric dipole moment experimental data $\overline{\theta }
\lesssim 10^{-11}$, and using $f_{a}\sim 10^{10}$ GeV, we find that the $N$
value, in order to keep PQ\ solution consistent must be $N\geq 10$. It means
that effective operators with $N<10$ must be forbidden by the $Z_{N}$
symmetry.

Before do that, we calculate the axion state. With the introduction of the
scalar singlet $\phi $, the scalar potential gains the following extra terms
\begin{equation}
V_{\phi \text{, extra}}=-\mu _{\phi }^{2}\phi ^{\dagger }\phi +\lambda
_{\phi }\left( \phi ^{\dagger }\phi \right) ^{2}+\lambda _{15}\left( \rho
^{\dagger }\rho \right) \left( \phi ^{\dagger }\phi \right) +\lambda
_{16}\left( \eta ^{\dagger }\eta \right) \left( \phi ^{\dagger }\phi \right)
+\lambda _{17}\left( \chi ^{\dagger }\chi \right) \left( \phi ^{\dagger
}\phi \right) \text{.}  \label{potencial singlete}
\end{equation}
Now, to calculate the eigenstate of the axion field, we write the fields as
\begin{align}
\rho & =\left(
\begin{array}{c}
\rho ^{+} \\
\frac{1}{\sqrt{2}}\left( V_{\rho ^{0}}+\text{Re\thinspace }\rho ^{0}+
i\text{Im\thinspace }\rho ^{0}\right) \\
\rho ^{++}
\end{array}
\right) \text{,}\qquad \eta =\left(
\begin{array}{c}
\frac{1}{\sqrt{2}}\left( V_{\eta ^{0}}+\text{Re\thinspace }\eta ^{0}+
i\text{Im\thinspace }\eta ^{0}\right) \\
\eta ^{-} \\
\frac{1}{\sqrt{2}}\left( V_{\eta _{1}^{0}}+\text{Re\thinspace }\eta
_{1}^{0}+i\text{Im\thinspace }\eta _{1}^{0}\right)
\end{array}
\right) \text{,}  \notag \\
\chi & =\left(
\begin{array}{c}
\frac{1}{\sqrt{2}}\left( V_{\chi ^{0}}+\text{Re\thinspace }\chi ^{0}+
i\text{Im\thinspace }\chi ^{0}\right) \\
\chi ^{-} \\
\frac{1}{\sqrt{2}}\left( V_{\chi _{1}^{0}}+\text{Re\thinspace }\chi
_{1}^{0}+i\text{Im\thinspace }\chi _{1}^{0}\right)
\end{array}
\right) \text{,}\qquad \phi =\frac{1}{\sqrt{2}}\left( V_{\phi }+\text{
Re\thinspace }\phi +i\text{Im\thinspace }\phi \right) \text{.}
\end{align}
The axion field must be isolated from the eight NG bosons that are absorbed
by the gauge bosons in the unitary gauge. This is fundamental to do a right
phenomenological study of the axion properties. By following standard
procedures, the axion field, $a\left( x\right) $, is determined to be
\begin{eqnarray}
a\left( x\right) &=&\frac{1}{f_{a}}\left[ \frac{V_{-}^{2}}{V_{\rho ^{0}}}
\text{Im\thinspace }\rho ^{0}-V_{\chi _{1}^{0}}\text{Im\thinspace }\eta
^{0}+V_{\chi ^{0}}\text{Im\thinspace }\eta _{1}^{0}+V_{\eta _{1}^{0}}
\text{Im\thinspace }\chi ^{0}\right.  \notag \\
&&\left. -V_{\eta ^{0}}\text{Im\thinspace }\chi _{1}^{0}-\left( \frac{
V_{-}^{2}}{V_{\rho ^{0}}^{2}}+\frac{V_{+}^{2}}{V_{-}^{2}}\right) V_{\phi }
\text{Im\thinspace }\phi \right] \text{,}
\end{eqnarray}
where
\begin{align}
V_{-}^{2}& \equiv V_{\chi ^{0}}V_{\eta _{1}^{0}}-V_{\chi _{1}^{0}}V_{\eta
^{0}}\text{,} \\
V_{+}^{2}& \equiv V_{\chi ^{0}}^{2}+V_{\chi _{1}^{0}}^{2}+V_{\eta
^{0}}^{2}+V_{\eta _{1}^{0}}^{2}\text{,}
\end{align}
and $f_{a}$ is the normalization constant given by
\begin{equation}
f_{a}\equiv \sqrt{\left( \frac{V_{-}^{2}}{V_{\rho ^{0}}}\right)
^{2}+V_{+}^{2}+\left( \frac{V_{-}^{2}}{V_{\rho ^{0}}^{2}}+\frac{V_{+}^{2}}{
V_{-}^{2}}\right) ^{2}V_{\phi }^{2}}\text{.}
\end{equation}
Note that in the limit $V_{\phi }\gg V_{\chi ^{0}}$, $V_{\chi _{1}^{0}}$, $
V_{\eta ^{0}}$, $V_{\eta _{1}^{0}}$
\begin{eqnarray}
a\left( x\right) &\simeq &-\text{Im\thinspace }\phi +\left( \frac{V_{-}^{2}}{
V_{\rho ^{0}}^{2}}+\frac{V_{+}^{2}}{V_{-}^{2}}\right) ^{-1}V_{\phi }^{-1}
\left[ \frac{V_{-}^{2}}{V_{\rho ^{0}}}\text{Im\thinspace }\rho ^{0}-V_{\chi
_{1}^{0}}\text{Im\thinspace }\eta ^{0}+V_{\chi ^{0}}\text{Im\thinspace }\eta
_{1}^{0}\right.  \notag \\
&&\left. +V_{\eta _{1}^{0}}\text{Im\thinspace }\chi ^{0}-V_{\eta ^{0}}\text{
Im\thinspace }\chi _{1}^{0}\right] \text{,}
\end{eqnarray}
i.e. the axion is primarily composed of the Im\thinspace $\phi $ field. As
it is well known, to do the invisible axion compatible with astrophysical
and cosmological considerations, the axion decay constant, $f_{a}$, must be
in the range $10^{9}$ GeV $\leq $ $f_{a}$ $\leq $ $10^{12}$ GeV.

Now, returning to the stabilization of the axion by the $Z_{N}$ symmetry,
let us put that in a short way. If the $Z_{N}$ symmetry that survives at low
energies was part of an \textquotedblleft anomalous\textquotedblright\ $
U\left( 1\right) _{A}$\ gauge symmetry, the $Z_{N}$ charges of the fermions
in the low energy theory must satisfy non-trivial conditions: The anomaly
coefficients for the full theory is given by the coefficients for the low
energy sector, in our case $A_{3C}\equiv \left[ SU\left( 3\right) _{C}\right]
^{2}U\left( 1\right) _{A}$ and $A_{3L}\equiv \left[ SU\left( 3\right) _{L}
\right] ^{2}U\left( 1\right) _{A}$, plus an integer multiple of $N/2$~\cite
{ibanez1993, babu2002}, i.e.
\begin{equation}
\frac{A_{3C}+pN/2}{k_{3C}}=\frac{A_{3L}+rN/2}{k_{3L}}=\delta _{\text{GS}}
\text{,}  \label{gs condiciones}
\end{equation}
with $p$ and $r$ being integers. The $k_{3C}$ and $k_{3L}$ are the levels of
the Kac-Moody algebra for the $SU\left( 3\right) _{C}$ and $SU\left(
3\right) _{L}$, respectively. In the present case these are positive
integers. Finally, the $\delta _{\text{GS}}$ is a constant that is not
specified by the low energy theory alone. Other anomalies such as $\left[
U\left( 1\right) _{A}\right] ^{3}$, $\left[ U\left( 1\right) _{A}\right]
^{2}U\left( 1\right) _{X}$ do not give useful low energy constraints because
these depend on some arbitraries choices concerning $U\left( 1\right) _{A}$~
\cite{banks1992}. This is why these do not appear in the Eq.~(\ref{gs
condiciones}). Now, to identify that anomalous $U\left( 1\right) _{A}$
symmetry, it is helpful to write it as a linear combination of the $U\left(
1\right) _{\text{PQ}}$ and the $U\left( 1\right) _{B}$ symmetries, i.e.
\begin{equation}
U\left( 1\right) _{A}=\alpha \left[ U\left( 1\right) _{\text{PQ}}+\beta
U\left( 1\right) _{B}\right] \text{,}
\end{equation}
where $\alpha $ is a normalization constant used to make the $U\left(
1\right) _{A}$-charges integer numbers. With the charges given in the Table
\ref{table 1}, it is straightforward to calculate the anomaly coefficients $
A_{3C}$ and $A_{3L}$,
\begin{equation}
A_{3C}=-\frac{3}{2}\alpha \text{, \ \ }A_{3L}=\left[ -\frac{9}{4}+\frac{3}{2}
\beta \right] \alpha \text{.}
\end{equation}
Thus, the $\beta $ parameter that satisfy the condition given in Eq.~(\ref
{gs condiciones}) is
\begin{equation}
\beta =\frac{1}{3}\left[ -3\frac{k_{3L}}{k_{3C}}+\frac{9}{2}+\frac{N}{\alpha
}\left( \frac{k_{3L}}{k_{3C}}p-r\right) \right] \text{.}
\end{equation}
Taking the simplest possibility for the parameters $k_{3C}$\ and $k_{3L}$,
i.e. $k_{3C}=$ $k_{3L}$, the parameter $\beta $ becomes
\begin{equation}
\beta =\frac{1}{3}\left[ \frac{3}{2}+\frac{N}{\alpha }\left( p-r\right)
\right] \text{.}
\end{equation}
Recalling that to stabilize the axion from the quantum gravity corrections
we need $N\geqslant 10$, we show two possible solutions with $N=10$ and $11$
. The corresponding charge assignment of these two discrete subgroups of the
$U\left( 1\right) _{A}$ symmetry are given in the Table \ref{table 3}. Also,
it is important to remember that those charges are defined mod $N$.
\begin{table}[th]
\caption{The charge assignments for $Z_{10}$ and $Z_{11}$ that stabilize the
axion, for $\protect\alpha=6 $.}
\label{table 3}\centering
\begin{tabular}{cccccccccc}
\hline\hline
& $\ Q_{\alpha L}$ \  & $\ Q_{3L}$ \  & $\ \left( u_{aR}\text{, }
u_{3R}^{\prime }\right) $ \  & $\ \left( d_{aR}\text{, }d_{\alpha R}^{\prime
}\right) $ \  & $\ \Psi _{aL}$ \  & $\ e_{aR}$ \  & $\ \rho $ \  & $\left(
\chi \text{, }\eta \right) $ \  & $\ \phi $ \  \\ \hline
$Z_{10}$ & $+5$ & $+7$ & $+1$ & $+1$ & $+7$ & $+1$ & $+6$ & $+6$ & $+2$ \\
$Z_{11}$ & $+6$ & $+7$ & $+1$ & $+1$ & $+8$ & $+2$ & $+6$ & $+6$ & $+4$ \\
\hline\hline
\end{tabular}
\end{table}

It can be explicitly verified that the charges in the Table \ref{table 3}
satisfy the Eq.~(\ref{gs condiciones}), as it should be, since $Z_{10}$ and $
Z_{11}$ are discrete subgroups of $U\left( 1\right) _{A}$, which is
anomaly-free by the Green-Schwarz mechanism.

\section{Conclusions}

In this paper we have shown a detailed and comprehensive study concerning
the implementation of the PQ symmetry into a 3-3-1 model in order to solve
the strong CP problem. We have considered a version of the 3-3-1 model in
which the scalar sector is minimal. In its original form this version has
only two scalar triplets ($\chi $,$\rho $) and it is found that the model
presents an automatic PQ-like symmetry. However, for this scalar content,
there is an $U(1)$ subgroup of $U(1)_{X}\otimes U(1)_{\text{PQ}}$ that
remains unbroken and hence no axion field, $a(x)$, arises. Therefore, the
strong CP problem can not be solved by the dynamical properties of the axion
field. However, as we have shown in the text, the problem can be solved due
to the appearance of three massless quark states. We show explicitly that
those massless quark states remain massless to all orders in perturbation
theory. This solution is disfavored since results from lattice and current
algebra do not point in that direction. When the model is slightly extended
by the addition of a third scalar triplet $\eta $, with the same quantum
numbers as $\chi $, we do not have massless quarks anymore but we can not
implement a PQ symmetry in a natural way. The trilinear term in the scalar
potential forbids this symmetry. We can resort to a $Z_{2}$ symmetry to
remove the trilinear term. In this case, we can define a PQ symmetry and an
axion field appears in the physical scalar spectrum. Unfortunately this
axion is visible since it is related to the $V_{\rho ^{0}}$ energy scale,
which is of the order of the electroweak scale. Therefore, the model must be
extended. We have succeeded in implementing a stable PQ mechanism by
introducing a $\phi $ scalar singlet and a $Z_{N}$ discrete gauge symmetry.
The introduction of the $\phi $ scalar makes the axion invisible provided $
V_{\phi }\gtrsim 10^{9}$ GeV, i.e. $a\left( x\right) \simeq \textrm{Im}\,\phi$. On
the another hand, the $Z_{N}$ protects the axion against quantum gravity
effects because both it is anomaly free, as it was shown by using a discrete
version of the Green-Schwarz mechanism, and it forbids all effective
operators of the form $\sim \phi ^{N}/M_{Pl}^{N-4}$, with $N<10$, which
could destabilize the PQ mechanism.

\acknowledgements B. L. S\'anchez--Vega was supported by CAPES.

\end{document}